# Dielectric relaxation and Charge trapping characteristics study in Germanium based MOS devices with HfO2/Dy2O3 gate stacks

# Dielectric relaxation and Charge trapping characteristics study in Germanium based MOS devices with $HfO_2/Dy_2O_3$ gate stacks

*M. Shahinur Rahman* [1-3,a] *, Member, IEEE, E .K. Evangelou* [3,b] *, Member, IEEE*

[1] GSI - Helmholtz Zentrum für Schwerionenforschung, D-64291 Darmstadt, Germany

[2] OncoRay-Medical Faculty, University of Technology- Dresden, D-01307 Dresden, Germany

[3] *Laboratory of Electronics-Telecommunications and Applications, Dept. of Physics, University of Ioannina, 45110-Ioannina, Greece.*

## Abstract

In the present work we investigate the dielectric relaxation effects and charge trapping characteristics of $HfO_2/Dy_2O_3$ gate stacks grown on Ge substrates. The MOS devices have been subjected to constant voltage stress (CVS) conditions at accumulation and show relaxation effects in the whole range of applied stress voltages. Applied voltage polarities as well as thickness dependence of the relaxation effects have been investigated. Charge trapping is negligible at low stress fields while at higher fields (>4MV/cm) it becomes significant. In addition, we give experimental evidence that in tandem with the dielectric relaxation effect another mechanism– the so-called Maxwell-Wagner instability– is present and affects the transient current during the application of a CVS pulse. This instability is also found to be field dependent thus resulting in a trapped charge which is negative at low stress fields but changes to positive at higher fields.

[a] Electronic mail: M.S.Rahman@gsi.de, Shahinur.Rahman@Oncoray.de
[b] Electronic mail: eevagel@uoi.gr
-Acknowledgement: We would like to thank Dr. Rosalind Perrin for reading the whole manuscript, valuable suggestions and corrections.

## 1. INTRODUCTION

As Germanium (Ge) offers higher mobility for electrons and holes when compared to Silicon (Si) it draws an extra attention in the semiconductor industry. In order to keep up with scaling requirements set by ITRS, gate dielectrics with higher permittivity ($\kappa$~25), like $HfO_2$, is used as a replacement of $SiO_2$ [1]. Germanium is highly reactive with $HfO_2$, which may lead to Ge diffusion into the $HfO_2$ dielectric [1]. One possible solution is the use of rare earth oxide dielectrics as interfacial buffer layers, which are "friendly" and can be directly deposited on Ge demonstrating better passivating and electrical properties [2]. $Dy_2O_3$ can efficiently eliminate Ge diffusion originating either from the substrate or from the interfacial layer, and also reduces the charge trapping effects while improving the equivalent oxide thickness (EOT) [3].

Another serious problem that arises when gate stacks of high- $\kappa$ dielectrics are used in MOS devices is that they all produce electrical instabilities in the corresponding devices. As a result, anomalous threshold voltage ($V_{TH}$) shifts [4] are observed. It also raises reliability concerns as it affects drive currents with time of operation. The position and spatial distribution of theses traps is also very important. Most of them lie in the bulk of the oxides and show dramatic transient effects in the drain current of MOSFET devices [5] or the leakage current of simple MOS capacitors [6]. In addition, when these traps lie close to the semiconductor-insulator interface they may respond to the applied ac signals thus leading to the concept of "Border traps" as introduced by Fleetwood et al. [ 3, 7] Moreover, all thin film dielectrics are definitely far from being considered as good insulators. While the use of relatively thicker high-k dielectrics - instead of thin $SiO_2$ - is a considerable improvement, these films still conduct dc current following one of the well-known current conduction mechanisms [8]. Therefore when a dc voltage is applied on the gate electrode of a MOS capacitor one of the following is likely to happen to the gate current $J_g$:

(i) The leakage current $J_g$ increases showing a charging capacitor behavior until - in a steady state condition – no more defects are available to trap carriers, *or*

(ii) The leakage current $J_g$ increases (Stress-induced leakage current, SILC), under bias condition (at high CVS), due to the creation of new neutral defects in the bulk of the oxides, *or*

(iii) When the defects lie close to the semiconductor or the metal gate electrode and/or their density and capture cross section is high, the fast initial charging leads to significant reduction of the field across the dielectric which is experimentally observed as a decay of $J_g$ with time.

High-$\kappa$ dielectrics are "trap-rich" materials [9] - [11] and charge trapping precludes accurate extraction of mobility of the devices [11]-[12]. The crucial concern is to understand why charge trapping takes place in gate-stack dielectrics. It has been widely accepted that the trapped charge resides in localized electronic states associated with structural defects [11], [13]-[18], pre-existing bulk defects [11], [19], dangling bonds at Ge-semiconductor /dielectric interface [20], oxygen vacancy and deviancies [21]. No matter what the origin is or whether they are bulk or interfacial defects they all give rise to transient gate currents with considerably high time constants.

Apart from these effects, which are commonly encountered in MOS devices with high-$\kappa$ oxide dielectrics; two more effects are likely to provide evidence of another source of unwanted transient currents. Relaxation effects and Maxwell-Wagner instabilities are both related to the multilayer structure of some gate dielectrics as will be explained in the following paragraphs.

In its simple form a MOS capacitor with a bilayer gate stack is usually studied with a thin (medium-$\kappa$) insulating layer in direct contact with the semiconductor surface and a thicker high-$\kappa$ oxide on top. The main reason for this structure is the experimentally proven and theoretically predicted fact, that the most interesting high-$\kappa$ oxides (e.g. $HfO_2$ or $ZrO_2$) for potential MOS devices produce very poor interfaces with a high density of electrically active defects. Thus a medium $\kappa$ buffer layer is utilized to suppress these interfacial defects. However, the existence of a high-$\kappa$ material introduces another undesirable effect: a relaxation current which follows the direction of the applied external voltage gradient, $dV_g/dt$ [22]-[24]. In general, relaxation in a solid involves the recovery of strain when the stress conditions change [24]. When an external field is applied across a film it separates the bound charges, thus resulting in polarization and a compensating internal field [25]. The physical nature of dielectric relaxation can be explained with a potential well model in terms of dipole orientation [26]. Dipoles, which are homogeneously distributed inside a material, are formed by localized defects and disorder due to a lack of crystallinity.

Recently, Jameson *et al.* [27] showed that the presence of a gate stack is itself one cause of charge trapping in the bulk of the dielectrics and/or at the interfaces between the two dielectrics and substrate-buffer layer. The problem has been recognized and was solved analytically many years ago [28]. It is due to the different insulating properties of the high-$\kappa$ layers in the gate stack, which results in different conductivity of each layer. Therefore, when a gate bias is applied to the stack charge drifts easily through the poorer insulating layer and accumulates at the interface of the two dielectrics. As a consequence the field across each insulator changes so that, after sufficient time has passed the same current density flows through both layers The effect which was described initially by Maxwell [29] himself, and later on by Wagner [30], is the so called "Maxwell-Wagner polarization", and causes current instabilities in voltage stressed dielectric stacks. This is due to charge accumulation at the interface of the two layers, which stimulates dielectric relaxation effects in each high-$\kappa$ layer. As dielectric relaxation is a continual buildup of polarization following the application of an electrical bias, it results in a transient displacement current through the dielectric. Therefore, this current instability due to "Maxwell-Wagner polarization" is also termed as "Maxwell-Wagner Instability (M-W)".

The afore-mentioned effects are already known to produce current instabilities in MOS devices containing various gate dielectrics. They both give a $J_g \sim t^{-n}$ behavior which is strongly voltage dependent [5], [22], [27]. Moreover, they are usually both present at the same time making the corresponding analysis a very complex task. The main subject of the present work is related analysis of the reliability issues of MOS devices comprising a dielectric gate stack. The studied devices grown on both *p*- and *n*-Ge substrates have been subjected to constant voltage stress conditions at accumulation. The aim of the present work is to identify Maxwell-Wagner instability and relaxation effects, as well as charge trapping at pre-existing bulk oxide defects, and to discuss potential reliability problems in future MOS devices.

## 2. EXPERIMENTAL

$Dy_2O_3/HfO_2$ oxide stacks were prepared by atomic oxygen beam deposition (MBD) on both *p*- and *n*-type Ge (100) substrates. Native oxide was desorbed *in-situ* under UHV conditions by heating the substrate to 360°C for 15 minutes until a (2x1) reconstruction appears in the (RHEED) pattern, indicating a clean (100) surface. Subsequently, the substrate was cooled down to 225°C where the oxide stacks were deposited. The surface was exposed to atomic O beams generated by an RF plasma source with the simultaneous e-beam evaporation of Dy/Hf at a rate of about ~0.15

Å/s. The same gate stacks ($HfO_2/Dy_2O_3$) of different compositions (nominal thicknesses), as well as single layer $Dy_2O_3$ were prepared on both *n*- and *p*-type Ge substrates for the present study, as shown in *Table I*. More details on the preparation and structural analysis of the devices can be found elsewhere [31]. Metal-insulator semiconductor capacitors were prepared by shadow mask and e-beam evaporation of 30 nm-thick Pt electrodes to define circular dots 200 μm in diameter. The back ohmic contact was made using a eutectic InGa alloy.

The devices were subjected to electrical stress under CVS conditions at accumulation [10]. Successive stress cycles of different time intervals and at different gate voltages were applied by means of a Keithley 617 source/ meter which was also measuring the corresponding current *vs* time ($J_g$-$t$) curves. After each stress cycle the gate bias was stopped in order to measure either the current-voltage ($J_g$-$V_g$) curves or the high frequency ($f$=100 kHz) capacitance-voltage ($C$-$V_g$) curve. This determined the flatband voltage shift ($\Delta V_{FB}$). The latter measurement was obtained by means of an Agilent 4284A LCR meter. For the *J-t* characteristics measurements, the capacitors were always biased at accumulation, and the absolute values of the current density and bias voltage were used in this study to avoid complexity. Fresh devices were used for each stress measurement with an area of $3.14\times10^{-4}$ $cm^2$. All the measurements were done in a dark box and at room temperature. The maximum change of temperature during the experiment was maintained within ± $0.2^0$C.

## 3. RESULTS AND DISCUSSION

### 3.1 Capacitance-voltage (*C*-V) characteristics under CVS

Typical $C$-$V_g$ curves of the MOS capacitors with a gate stack dielectrics at low and moderate bias are illustrated in Fig.1 (a) and (b) respectively. In order to measure the trapped oxide charges immediately after stopping the stress pulse, the curves were obtained from accumulation to inversion and backwards at a gate voltage sweep rate of 100 $mVs^{-1}$. This corresponded to switching times of ~40 s over the portion of the curve showing hysteresis. Ten successive CVS cycles of 500 s each were applied, and for sake of clarity, the curves of the fresh device and after the 10th stress are shown in the figures. Nevertheless, the important electrical properties of the capacitors (for example the EOT or the density of interface states do not show substantial differences from the $C$-$V_g$ acquired in the opposite way which is typically used (i.e. from inversion to accumulation and backwards). The hysteresis of the $C$-$V_g$ curves was rather large (around 400 mV at midgap) and a large density of slow interface traps is evident even at ac signal frequencies as high as 100 kHz. The corresponding current-voltage ($J_g$-$V_g$) curves show very small leakage currents (around 15 $nA/cm^2$ @ ±1 V-$V_{FB}$) [31].

The interesting result from the analysis of the high frequency $C$-$V_g$ curves of sample P4 (Fig. 1(a) and (b)) is that when the applied stress voltage is rather low, i.e. $Vg$= -2 V ($E_{HfO2}$=1.0 MV/cm, $E_{Dy2O3}$=1.9 MV/cm), the trapped charge in the oxide is negative (i.e. $\Delta V_{FB}$ shift is positive). However, at moderate stress voltages i.e. $Vg$= -3V ($E_{HfO2}$=1.8MV/cm, $E_{Dy2O3}$=3.1MV/cm) the observed negative shift of the $C$-$V_g$ curves indicates positive charge trapping. Similar results have been observed on all other gate stacks (see *Table I*) and there are two possible explanations of the observed phenomenon: Firstly, as the gate voltage during the stress pulse is always negative, electrons are injected into the dielectrics from the metal. At low voltages these electrons are trapped in pre-existing defects and the fields across each dielectric are not high enough for these electrons to escape towards the *p*-Ge substrate. At higher stress voltages the situation is different as holes are injected from the p-Ge substrate into the oxide, thus resulting in the positive charge trapping. Also, at the same time, a significant amount of new positive defects are created in the bulk of the oxides.

A different approach is to take into consideration the fact that, because the conductivities of $HfO_2$ and $Dy_2O_3$ thin films depend differently on the applied field, either layer can have the higher conductivity depending on the choice of gate voltage. Frohman-Bentchkowsky and Lenzlinger [28] caused the sign of the trapped charge to switch by varying the gate voltage of similar (gate stack) structures. This effect was predicted from the independent measurements of the conductivities of the two layers [27]-[28]. Similar changes of sign might have already been observed in $HfO_2/SiO_2$ gate stacks [28]. Furthermore, in previous work [10], we observed and reported the same effect on MOS devices with $CeO_2$ as the gate dielectric.

In order to check which of the above mechanisms is responsible for the observed $V_{FB}$ shifts, the transient currents which are present during the application of a CVS pulse were measured. The corresponding analysis is presented in the following paragraphs.

### 3.2 Voltage dependence of dielectric relaxation:

#### 3.2.1. Substrate dependence of $J_g$-$t$ curves

As the direction of the leakage and relaxation currents depend on the polarities of $V$ and $dV$ respectively, their magnitude can be either additive or subtractive. The directions of these two currents through the high-$\kappa$ gate stack of a *p*-Ge based device are illustrated in Fig. 2. When a negative gate voltage pulse is applied, the device is driven in accumulation and the relevant leakage current is negative. At the same time as $dV<0$ the magnitude of the relaxation current is also negative.

In order to study the current transient characteristics of both *n*- and *p*- Ge based MOS devices we applied different CVS bias (from |1| to |5|V) on samples # P2 and N1, and the corresponding fields are given in *Table II*. The corresponding current densities as

a function of stress time ($J_g$-$t$) curves are shown in Figs. 3(a) and (b). Interestingly, on *p*-Ge based devices and low CVS conditions, a decaying current which follows a power law ($t^{-n}$) is observed [see Fig. 3(a)]. For the gate stacks grown on *n*-type substrates, this current decay is never traceable even at very low CVS conditions [see Fig. 3(b)]. On the contrary, at higher stress voltages and on both type of substrates, we do not notice dielectric relaxation because of the dominating charge trapping mechanism that will be discussed in a later section. Soft Breakdown (SBD) and Hard Breakdown (HBD) events have also been detected at higher fields and/or prolonged time stress [Fig. 3(a)].

In order to better understand which mechanism is responsible for the change of direction of the $\Delta V_{FB}$ shift with the applied gate voltage the transient response of the current during the application of the stress pulse was monitored in more detail [see Fig. 1(a) and (b)]. Fig. 4 illustrates the current density $J_g$ versus stress time $t$ curves after the application of relatively moderate stress voltages on *p*-Ge based devices (in the form of train pulses). During the CVS measurement we recorded the $J_g$-$t$ curves after the application of ten consecutive stress pulses, each one having duration of 500s while the gate voltage was kept constant [10]. Between the voltage pulses, $J_g$-$V_g$ curves at accumulation were also acquired. In Fig. 4 only the first and last curves are plotted for the sake of clarity. The decay of $J_g$ follows a $t^{-n}$ law with $n$ values varying smoothly from 0.73 to almost unity. Additionally the $n$ values increase continuously in every new stress cycle reaching a value of 0.91 after 10 successive cycles. The fact that the initial n value is far from unity indicates that a Maxwell–Wagner instability (following the terminology used in [27]) is likely to be present together with the usual dielectric relaxation of the high-$\kappa$ dielectrics. In the latter case the relaxation current decays with time following the Curie–von Schweidler relaxation law(C-S) [24]:

$$J_e = C \cdot t^{-n}, \qquad (1)$$

where $J_e$ is the relaxation current density (A/cm$^2$). $C = P.\alpha$ where $P$ is the total polarization or surface charge density (V·nF/cm$^2$), $\alpha$ is a constant in seconds and $n$ is a real number close to unity. The gradual increase of $n$ could be attributed to the fact that the Maxwell–Wagner instability becomes less important after each stress cycle. The relevant $J_e$ values decrease so that after 10 consecutive cycles the dielectric relaxation current dominates. One possible explanation for this effect is the gradual change of the conductivities of the two dielectric layers, due to charge trapping on preexisting bulk oxide defects.

### 3.2.2. Thickness dependence of dielectric relaxation

Fig. 5 shows that the relaxation current increases linearly with increasing gate bias for three different gate stack configurations (samples #P2, #P3, #P4). The current measured at $t$ = 3 s ($J_{g=3s}$), after setting the stress pulse, is used as a measure of the amplitude of the relaxation current. From (1), the magnitude of the relaxation current is directly proportional to the applied voltage across the dielectric. Therefore a linear $J_g - V_g$ plot indicates the presence of relaxation currents rather than any other transient mechanisms. It should be noted here that due to rise time limitations of the measuring instrument the $J_g$ data acquired for $t$ < 1s are not taken into account. Jameson *et al.* [27], Reisinger *et al.* [32], and Luo *et al.* [22] observed similar current decays on Si-based devices which were attributed to the relaxation of the dielectric material, while Xu *et al.* [33], and Bachhofer *et al.* [34] explained these effect by charge trapping-detrapping within the gate dielectrics.

In order to explain which of the above models apply to our results, relaxation current densities ($J_e$) at 3s as a function of the electric field across (a) $HfO_2$ and (b) $Dy_2O_3$ are plotted in Fig. 6 (a) and (b). From the figures there is a clear indication of the thickness independence of the relaxation current. This is expected as the amplitude of polarization is controlled by the electric field across the dielectric materials. As a result, the corresponding current should be identical when induced by the same electric field and independent of the film thickness variation [24]. Similar results have been reported by Reisinger *et al* in BSTO [32] films. This thickness independence is consistent with the normal dielectric material polarization model [24], [35], and can not be explained by charge trapping and detrapping mechanisms [33]. As $Vg$ is negative, the electrons are injected from the gate electrode. This means that the calculations of the initial electric fields across the $HfO_2$ and $Dy_2O_3$ films are very important factors. The field across each of the layers of the gate stack can be calculated as [36]:

$$E_{HfO_2} = \frac{V}{(\kappa_1/\kappa_2)d_2 + d_1} \qquad (2)$$

$$E_{Dy_2O_3} = \frac{V}{(\kappa_2/\kappa_1)d_1 + d_2} \qquad (3)$$

where, $V = V_g - V_{FB} - \Psi_s$ is the voltage applied to the gate dielectric stack, $V_{FB}$ is the flatband voltage, and $\Psi_s$ is the initial surface potential of Ge. $d_{1,2}$ *is* the thickness of the high-$\kappa$ ($HfO_2$) or the interfacial ($Dy_2O_3$) layer respectively, $\kappa_1$ and $\kappa_2$ being their dielectric constants respectively. All field values in the present work were calculated using equations (2) and (3). It should be pointed out that the calculation of the initial electric field in the high-$\kappa$ film, $HfO_2$ (2), as well as the initial field across the $Dy_2O_3$ (3), is only an estimation of the magnitudes, and will be discussed in the next section.

### 3.3 Correlation of Dielectric Relaxation and Maxwell-Wagner Instability

As has been discussed earlier, because of the bilayer structure some charge is accumulated at the interface between the two dielectrics due to the "Maxwell-Wagner

Instability" [27]. In addition, if one tries to fit the experimental $J_g$-$t$ data by means of the Curie–von Schweidler law alone, the calculated values of $n$ are less than unity ($n$ ~ 0.73). However after successive CVS cycles (i.e. continuous charge injection) this value of $n$ tends to unity ($n$ = 0.91), which could be explained if one assumes that the relaxation effects and the "Maxwell-Wagner Instability (M-W)" act simultaneously. According to the potential well model [26], the current due to relaxation from a single dielectric layer is

$$J_g = 2\sigma_0 \frac{V}{d}\left(3 + \ln\frac{t}{t_{0,1}}\right)\frac{t_{0,1}}{t} \quad t>t_o \qquad (4a)$$

where V is the applied external bias, d is the thickness of the dielectric while $t_o$ and $\sigma_0$ are material constants. In general $t_o$ is expected to be of the order of picoseconds while $\sigma_0$ is not related to the dc conductivity of the insulating oxide layer. In the case of a gate stack configuration, where the two dielectrics are perfect insulators, the field across each dielectric will be different than the simple V/d factor of (4a).

However, the first dielectric ($k_1$ in Fig. 2), which is deposited on top of the semiconductor surface, is usually very thin and mainly amorphous. It is then reasonable to assume that it does not contribute to the relaxation current. However, it does modify the field across the top dielectric and a M-W factor is introduced. Therefore, the relaxation current due to these combined effects can be expressed as [27]:

$$J_g = 2E_{HfO_2}\sigma_{0,1}\left(3 + \ln\frac{t}{t_{0,1}}\right)\frac{t_{0,1}}{t}, \quad t>t_{0,1}. \qquad (4b)$$

where $$E_{HfO_2} = \frac{V\kappa_2}{d_1\kappa_2 + d_2\kappa_1}$$

is the field across the high-k material ($HfO_2$ in this case) while $\sigma_{0,1}$ and $t_{0,1}$ are the material constants which set the scale of current and time respectively and all other terms have been mentioned before in (2)-(3).

At this point it is interesting to notice that (4b) could only be utilized for the present gate stacks under the following assumptions:

i. the REO buffer layer is thin and amorphous so that the corresponding relaxation effects are suppressed. Otherwise a second term (which accounts for the relaxation in the buffer layer) must be added in (4b),

ii. Eq. (4b) could only fit the experimental *J-t* data for a short time interval (usually <100 s) as it does not take into account leakage current effects, and

iii. the field, $E_{HfO_2}$, across $HfO_2$ may differ from $V\kappa_2/(d_1\kappa_2+d_2\kappa_1)$ by an amount depending on the magnitude of the interfacial charge σ as explained in detail in [26]. One way to obtain accurate interfacial charge (σ) values is the use of the correct conductivities $J_1(E_1)$ and $J_2(E_2)$. Without knowledge of the conductivity of each dielectric layer, $V\kappa_2/(d_1\kappa_2+d_2\kappa_1)$, is only an approximation, which is based on the fact that as the relevant change of the field across each dielectric is small. The conductivity could be approximated by a linear (i.e. ohmic) behavior.

The above mentioned prerequisites could not be met in all samples and stress voltages used in this study. So, the model was only used to explain the deviation from the Curie von Schweidler ($J{\sim}t^{-1}$) law.

Fig. 7 shows the current density as a function of stress time at different low gate voltages. It should be mentioned here that the use of gate voltage *Vg* as the changing parameter was chosen in many plots in the present work. This was done as the use of the corresponding fields (by means of (2)-(3)) turns out to be very complicated. We fit the experimental data for two different thicknesses of $HfO_2/Dy_2O_3$ gate stacks and two different $V_g$ values using (4b). The thickness of each layer is obtained from independent measurements, while *V, J* and *t* are derived from experimental data. Therefore, in order to find the values of the free running parameters $\sigma_{0,1}$, $t_{0,1}$ and $\kappa_{1,2}$, two different sets of experimental *J-t* data were acquired after application of different $V_g$ voltages on the same sample. Fitting (4b) to the experimental data the relevant parameters have been calculated as $\kappa_1 = 20$, $\kappa_2 = 13$, $\sigma_{0,1} = 2\sim3\times10^{-5}$ A/cm$^2$ and $t_{0,1}$=2.1×10$^{-11}$s respectively. It should be noted here that an accurate solution of the four unknown parameters of (4) needs a set of four $J = f(V_g,t)$ equations. However, the separation of parameters in (4) and the initial guess values for $\sigma_{0,1}$ and $t_{0,1}$ obtained from similar analyses in ref [27] was proved to be good enough for the excellent fit shown in Figs. 7(a) and (b). The addition of two more $J= f(V_g,t)$ experimental curves does not alter the obtained values significantly.

Comparable $\kappa$-values of $HfO_2$ [37] and $Dy_2O_3$ [38] have been confirmed by means of high frequency *C-V* measurements [31] on similar samples (see *Table III*). Therefore, it should be emphasized here that the $\kappa$-values obtained after fitting (4b) to the experimental data is another measure of the validity of the model described by (4b) under the relevant assumptions. Furthermore, in an attempt to fit a simple relaxation power law ($J_e \sim t^{-n}$) to the experimental data of moderate to high applied $V_g$ values [see Fig. 4], the obtained exponent value deviated considerably from unity. In addition, the exponent *n* was never the same during the first stress cycle when slightly different stress voltages were applied to the same sample. It was then reasonable to assume that the current decay was not due to relaxation effects alone. On the contrary, when the applied CVS values were lower than 1.5 V, the relaxation effects dominate, and the use of (4b) explains the deviation of the exponent *n* from unity.

In order to show the validity of (4b), for the case of a gate stack configuration, one can check it against a set of various thicknesses of the two oxides. After fitting the experimental data using (1) (when bias is applied to the MOS capacitors) corresponding pre-exponential factors *C* as a function of gate bias is illustrated in Fig. 8(a). The variation of thickness, both for the high-$\kappa$ and interfacial layers, results in notably different *C* lines as shown in Fig.8 (a). However, when the time independent coefficients of (1) and (4b) are considered, the coefficient C is equal to $C = 2E_{HfO_2}\sigma_{0,1}$, while the time dependent terms of both equations are practically indistinguishable. The coefficient, *C,* versus the field across the high-$\kappa$ dielectric ($E_{HfO_2}$) is plotted in Fig.8 (b). The experimental data in this case lie one on top of another. This figure illustrates that this scaling holds true, meaning that the thickness dependence of (4b) is correct for the case where the thickness of the interfacial layer varies (2-5nm) while that of the high-$\kappa$ layer is held fixed. Moreover, the thickness dependence of (4b) is also correct when the thickness of the interfacial layer is held fixed while that of $HfO_2$ varies (5-8 nm) [see Fig. 8(b) 'insert']. Jameson *et al.* [27] reported similar result for $HfO_2/SiO_2$ based devices on *p*-Si substrates.

### 3.4 Dielectric relaxation and Charge trapping characteristics at higher stress voltages

The application of higher stress voltages on the same MOS devices results in quite different transient characteristics of the corresponding $J_g$-*t* curves. As illustrated in Fig.9 (a), upon application of moderate to high stress voltages, i.e. *Vg* = -4.8V ($E_{Dy2O3}$=4.8MV/cm) on the single $Dy_2O_3$ devices the relaxation effects disappear. The transient current behavior is now governed by charge trapping at preexisting bulk oxide defects. In contrast, application of moderate stress voltages, i.e. Vg = -3.0V ($E_{HfO2}$ = 3.3MV/cm, $E_{Dy2O3}$ = 5.9MV/cm) on capacitors with the $HfO_2/Dy_2O_3$ stack (Fig. 9b, sample #P2), show the coexistence of two different mechanisms separated only by the different time scales of each one. Therefore, during the first 32 seconds after the application of the pulse, the current density $J_g$ decreases with time due to the relaxation mechanisms. This follows a $t^{-n}$ law with n values as low as 0.6. At the same time the magnitude of the leakage current that flows through the dielectrics is 2-3 orders of magnitude higher than in the case of low stress voltages [see for example Figs. 4 and 7(a)]. Therefore the charge trapping effects become more significant and the $J_e$ values start to increase following a model originally proposed by Nigam *et al.* [39] to explain charge trapping in MOS devices with thin gate stack dielectrics [40]

$$J_g - J_o = N^+(V_g) \cdot \left[1 - e^{-\frac{t}{\tau}}\right] + \alpha \cdot t^{\nu} \tag{5}$$

with $N^+(V_g)$ being the saturation value of positive charge trapping, $\tau$ the trapping time constant and $\alpha$ and $\nu$ the SILC related parameters, and $J_o$ the first value of current density. The first term in (5) represents an exponentially saturating charge build-up on pre-existing oxide defects, while the second term represents the increase due to SILC-generation.

According to this equation (5), the transient behavior of $J_g$ with time, for sample #P1 see Fig. 9(a) could be explained by taking into consideration both trapping on preexisting bulk oxide defects (with a characteristic time constant $\tau \sim 32$s) and creation of new defects due to electrical stressing (which follow a power law $J_g \sim t^{\nu}$, as in (5)). However for sample #P2, only charge trapping was considered for best fitting of the experimental data [see Fig. 9(b)]. In addition, the time constant $\tau$ is one order of magnitude greater ($\tau \sim 260$s) for that device than for the structure containing only $Dy_2O_3$. This is an interesting result as it shows that there are different types of defects in the two oxides. Furthermore, the overall better insulating properties of $HfO_2$ are confirmed. Sample #P2, although stressed at slightly higher electric fields, shows negligible rate of creation of new defects. Similar effects have been observed for the other devices with bilayer dielectrics studied in the present work, as illustrated in Fig. 9(c). In this figure the existence of both soft (SBD) and hard (HBD) breakdown effects is clearly demonstrated for moderate to high CVS conditions.

## 4. CONCLUSION

Charge trapping and relaxation characteristics of Pt/$HfO_2$/$Dy_2O_3$ /Ge gate stacks were studied by means of CVS measurements. At low applied stress voltages two independent electrical instabilities were observed namely, the Maxwell–Wagner instability and dielectric relaxation. While both effects were present simultaneously, the increase of the applied voltage and/or the repetition of the stress cycles led to a change of the relative magnitude of each one separately. Another aspect of the studied structures worth noting is that, because of the different effects dominating at low to medium or high applied fields, the sign of the trapped charge switched from positive to negative, an effect that has been rarely reported for high-$\kappa$ gate stacks. Finally, at moderate to high stress fields, the dominant process is charge trapping and creation of new defects (SILC). The analysis of the transient behavior of the current density in this case revealed the existence of two different trapping centers in the two dielectrics at least in terms of the relevant capture cross sections.

**TABLE CAPTIONS**

***Table I*****:** Summary of gate stack dielectrics deposited at different nominal thickness and types of Ge substrates

***Table II*****:** Calculation of the applied gate voltages and the corresponding electric fields according to (2 and 3) for samples (a) #P2 and (b) #N1 at time t=0s.

***Table III*****:** Dielectric constant (*κ*- values) and equivalent oxide thickness (EOT) values of $HfO_2, Dy_2O_3$ and $HfO_2 / Dy_2O_3$ gate stacks, from fit of eq. (4b) to the experimental data and after a recent study (*ref. 31*) of high frequency C-V curves.

## FIGURE CAPTIONS

**Fig 1 (a, b, c):** High frequency $C$-$V_g$ ($f$ =100 kHz) curves on fresh and stressed devices of sample P4. Only the curve after the application of ten consecutive CVS cycles (500s each) is plotted for clarity. Stress voltage is low in (a) and moderate in (b). Positive $V_{FB}$ shifts in (a) indicate trapping of electron in the bulk of the oxides while negative $V_{FB}$ shifts in (b) indicate creation of positively charged defects.

**Fig. 2:** Schematic diagram of the leakage and relaxation currents of an MOS device biased at accumulation.

**Fig.3 (a,b)** Current density as a function of stress time curves ($J_g$-$t$) at different CVS conditions of a gate stack grown on (a) *p*- and (b) *n*-type Germanium substrates (samples P2 and N1 respectively). The corresponding fields across each dielectric are given in *Table II*

**Fig. 4** The figure illustrates absolute values of current density ($J_g$) as a function of stress time, $t$. The transient current behavior during the application of the first and the tenth stress pulses is shown for clarity. The change of slope is rather smooth for the corresponding curves obtained during the application of intermediate CVS pulses (i.e. 2nd to 9th). The applied stress field is low for this gate stack configuration (sample #P2). Solid lines represent the Curie–von Schweidler relaxation $t^{-n}$ fit to the experimental data.

**Fig. 5:** Gate relaxation current, measured 3 seconds after setting the stress pulse for three different gate stacks ($HfO_2/Dy_2O_3/p$-Ge), as a function of applied CVS bias V The gate stacks was grown on *p*-type Ge substrates and the applied CVS bias was negative, that is, at accumulation. The relaxation current changes linearly with $V_g$ and is thickness dependent.

**Fig. 6 (a, b)** Gate relaxation current measured at 3s as a function of (a) $HfO_2$ high-$\kappa$, and (b) $Dy_2O_3$ interfacial layer electric fields in p-substrates MOS-capacitors. Relaxation current is thickness independent on $HfO_2$ or $Dy_2O_3$ electric fields that anticipates polarization model, and is incompatible to the charge trapping-detrapping model. Solid lines are simply a guide to the eye.

**Fig. 7** Gate current as a function of stress time of two different thicknesses gate stacks (#P2, #P3). Solid lines are best fit to (4b) and indicate that the combined effect of relaxation and Maxwell-Wagner instabilities, better describes the observed current decay.

**Fig. 8 (a, b)** Experimental results of the dielectric relaxation current in high-k $Pt/HfO_2/Dy_2O_3$/p-Ge gate stacks. (a) Coefficient *C* in fits of $C/t^n$ to the dielectric relaxation current of gate stack capacitors biased into accumulation (Vg negative). (b) Same data as in (a), but with the horizontal axis scaled in $E_{HfO_2} = V\kappa_2 / d_1\kappa_2 + d_2\kappa_1$ according to (4), making the data to collapse onto a single line. *C* vs $E_{Dy2O3}$ data are shown as an insert in Fig. 8(b). Solid lines are simply a guide to the eye.

**Fig. 9 (a, b, c)** $|J_g|$ *vs t* curves are shown, when the applied gate voltages are rather high, so that the corresponding fields are moderate for all samples #P1, #P2, and #P4 (a, b and c respectively ). Solid lines are best fit to the experimental data according to (5).

| Samples reference | Structures (compositions) |
|---|---|
| N1 | Pt/$HfO_2$(10nm) /$Dy_2O_3$(1nm)/*n*-Ge |
| P1 | Pt/$Dy_2O_3$(10nm)/*p*-Ge |
| P2 | Pt/$HfO_2$(5nm) /$Dy_2O_3$(2nm)/*p*-Ge |
| P3 | Pt/$HfO_2$(8nm) /$Dy_2O_3$(2nm)/*p*-Ge |
| P4 | Pt/$HfO_2$(5nm) /$Dy_2O_3$(5nm)/*p*-Ge |

| Fig. 3(a) sample #P2 | | | Fig. 3(b) sample #N1 | | |
|---|---|---|---|---|---|
| Gate voltage (V) | Corresponding fields across gate stacks (MV/cm) | | Gate voltage (V) | Corresponding fields across gate stacks (MV/cm) | |
| $-V_g$ | $E_{HfO2}$ | $E_{Dy2O3}$ | $V_g$ | $E_{HfO2}$ | $E_{Dy2O3}$ |
| 0.7 | 0.6 | 1.1 | 1.7 | 1.4 | 2.5 |
| 1.4 | 1.4 | 2.6 | 2.2 | 1.9 | 3.3 |
| 2.1 | 2.3 | 4.0 | 2.8 | 2.3 | 4.2 |
| 2.8 | 3.1 | 5.8 | 3.3 | 2.8 | 5.0 |
| 3.5 | 3.9 | 6.9 | 3.9 | 3.3 | 5.8 |
| 4.1 | 4.5 | 8.1 | 4.4 | 3.7 | 6.7 |
| 4.2 | 4.7 | 8.4 | 4.8 | 4.1 | 7.3 |
| | | | 5.3 | 4.5 | 8.0 |

| $HfO_2/Dy_2O_3$ **gate stacks** | | | | | **experimental results for single $HfO_2$ or $Dy_2O_3$ layers** | | **EOT** | | | |
|---|---|---|---|---|---|---|---|---|---|---|
| **present study** | | **C-V measurement *(ref.31)*** | | | | | $HfO_2/Dy_2O_3$ **gate stacks *(ref.31)*** | | | |
| $K_{HfO2}$ | $K_{Dy2O3}$ | $K_{HfO2}$ | $K_{Dy2O3}$ | *Effective K (gate stacks)* | $K_{HfO2}$ (*ref*.37) | $K_{Dy2O3}$ (*ref*.38) | P1 | P2 | P3 | P4 |
| 20 | 13 | 23~25 | 13~14 | 16~18 | 20~25 | 12~14 | 2.68 | 1.93 | 2.29 | 2.77 |

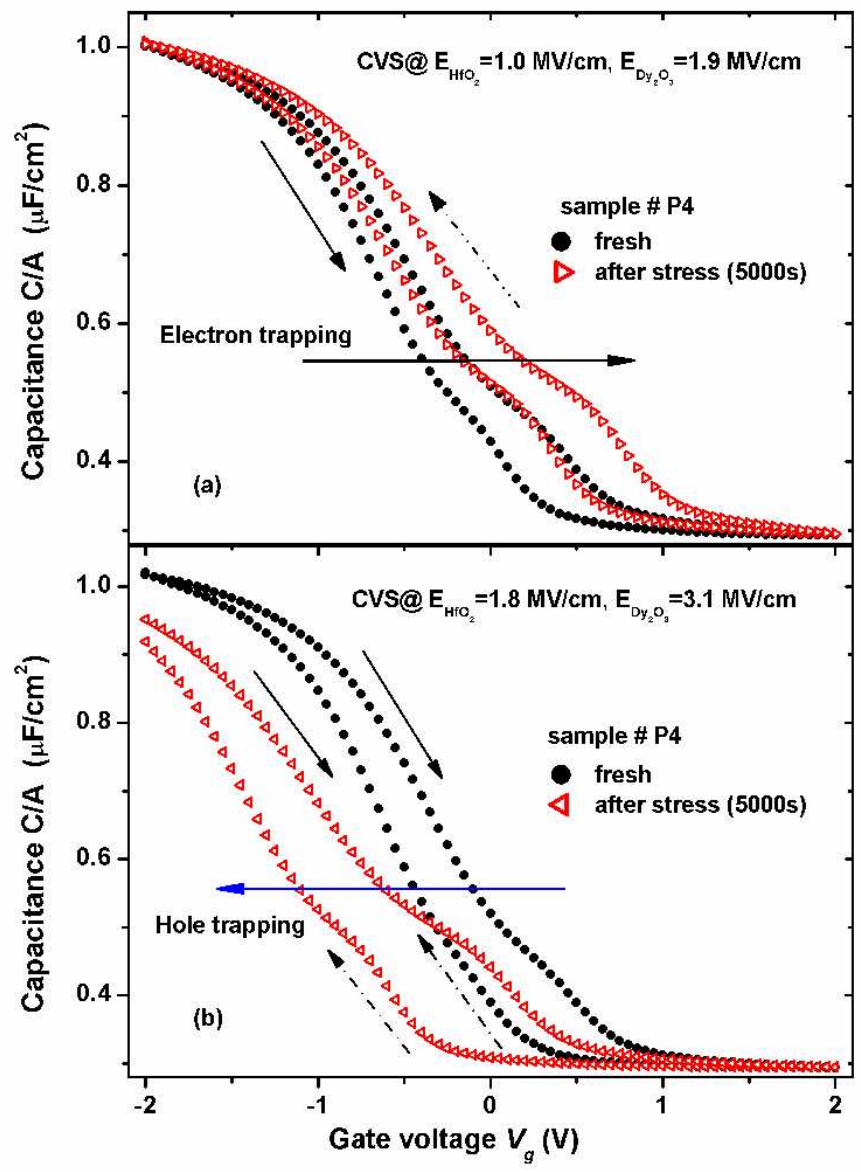

Fig1

289x202mm (300 x 300 DPI)

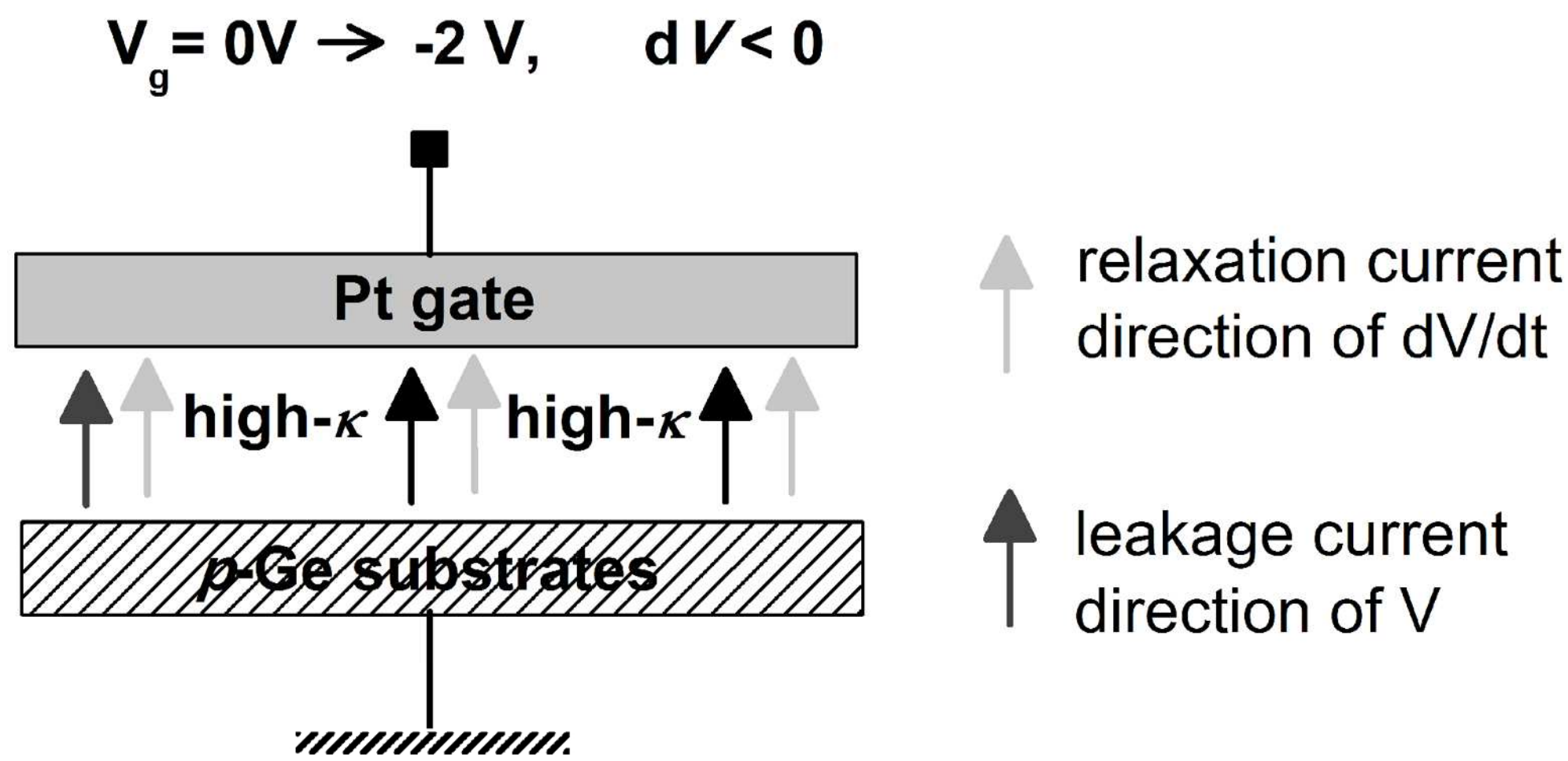

Fig. 2

82x52mm (600 x 600 DPI)

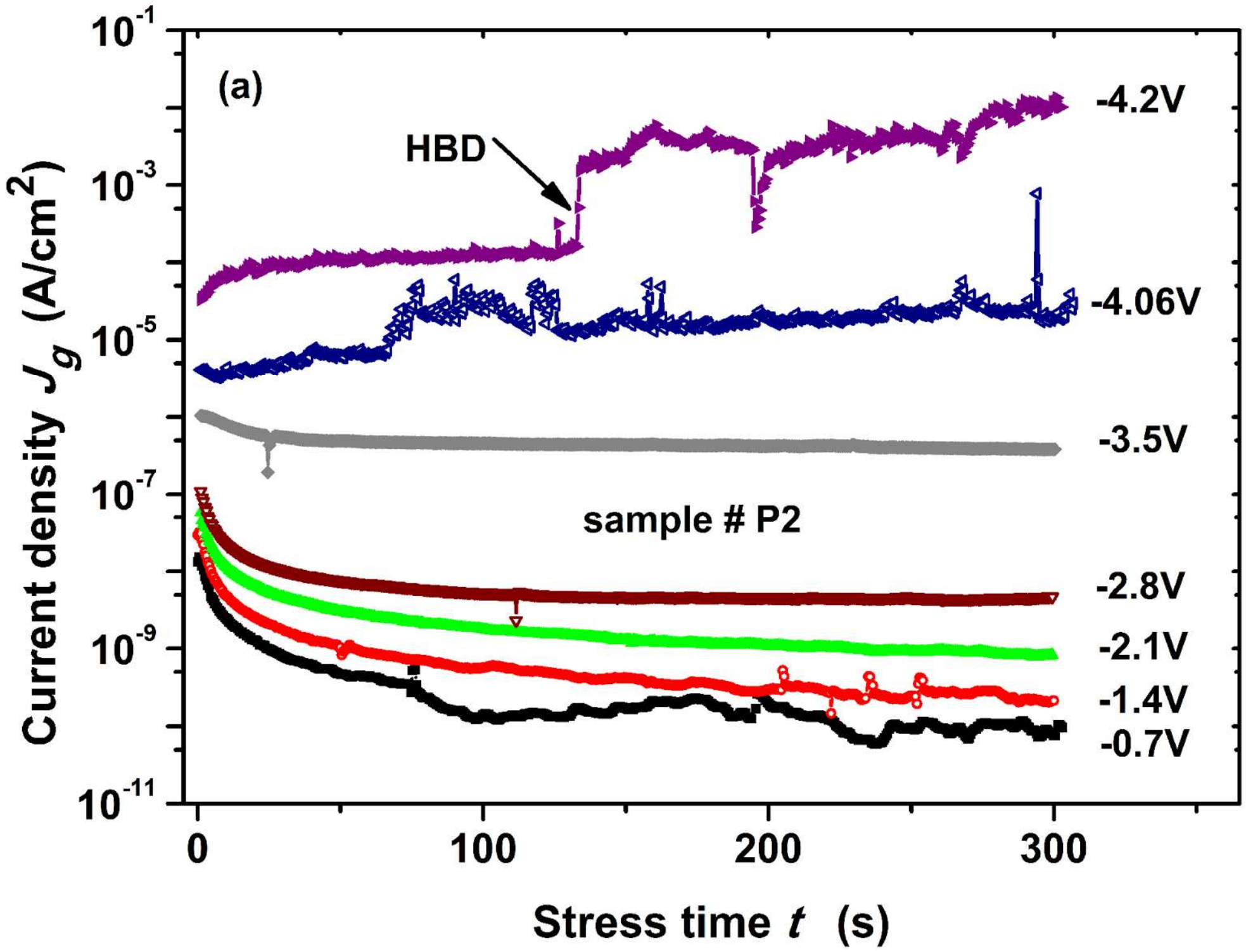

Fig. 3(a)

82x66mm (600 x 600 DPI)

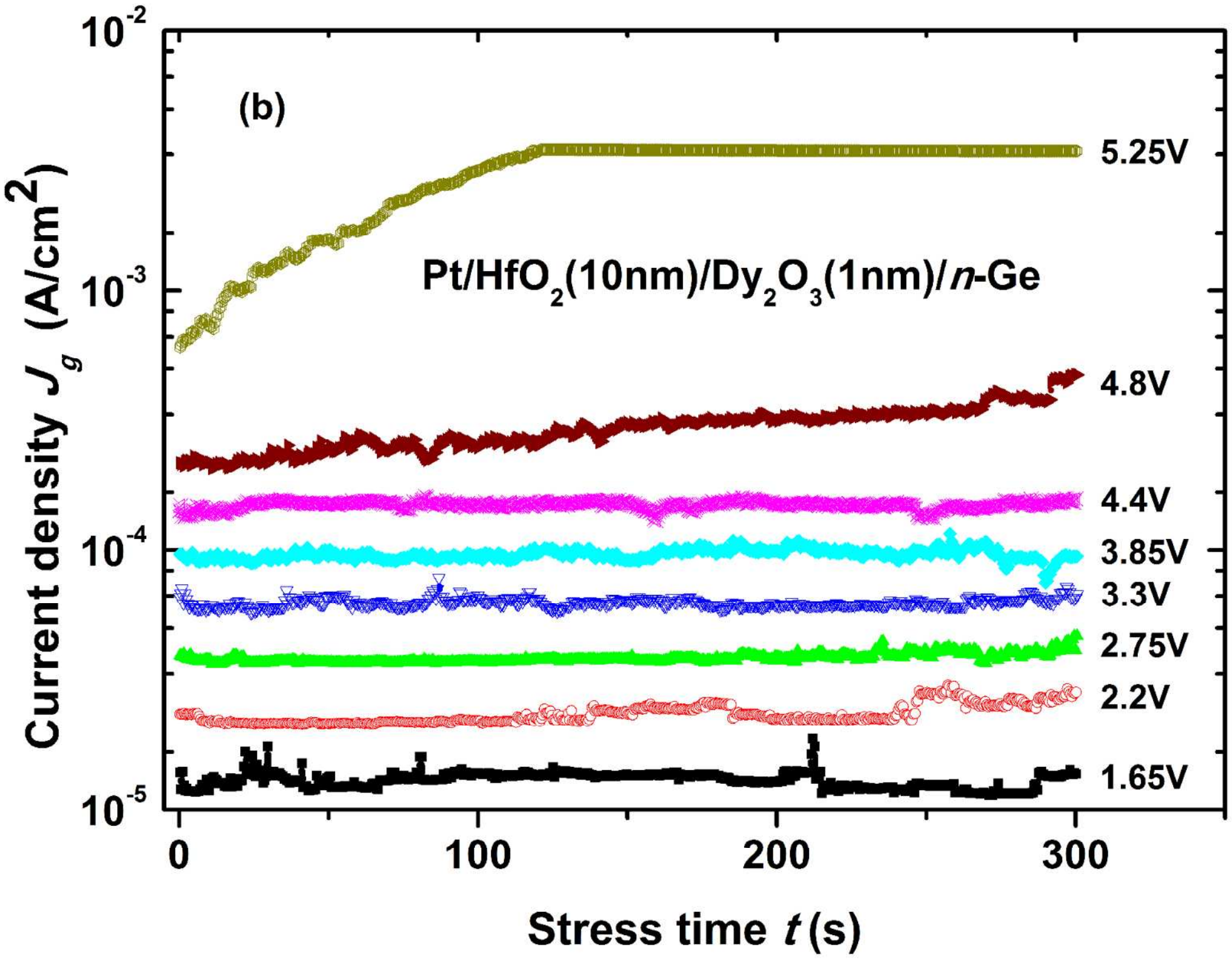

82x64mm (600 x 600 DPI)

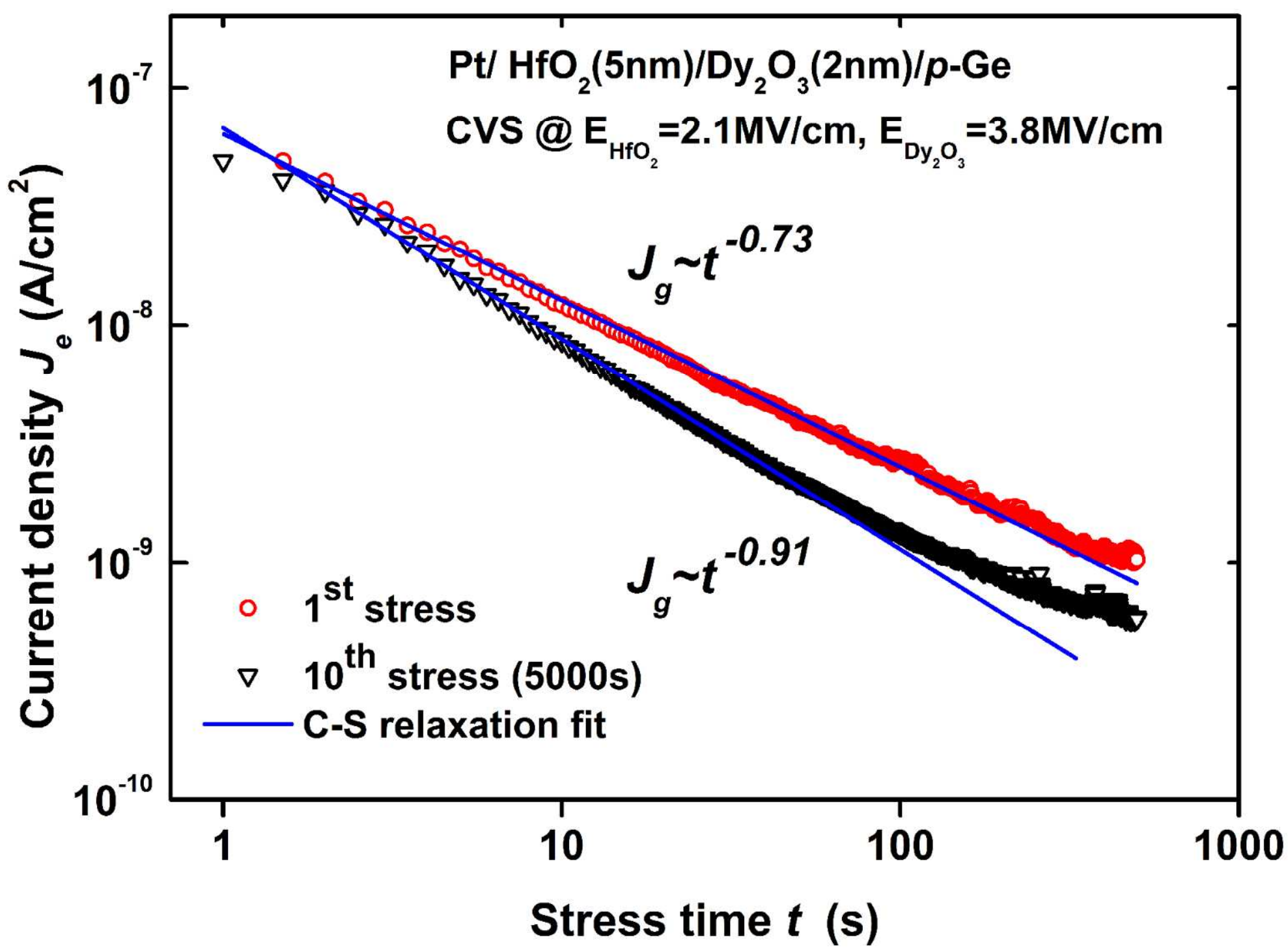

Fig.4

82x63mm (600 x 600 DPI)

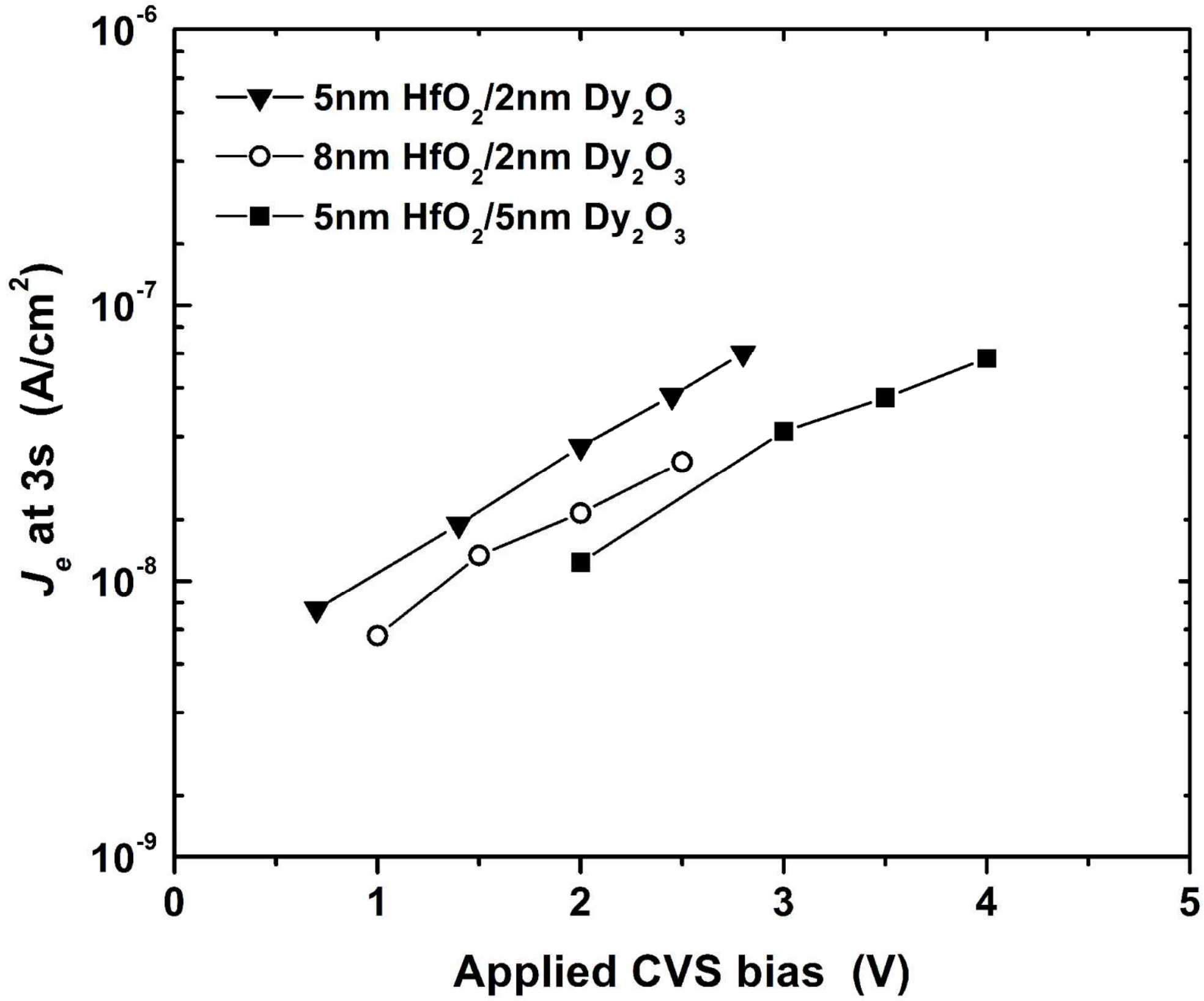

Fig5

82x72mm (600 x 600 DPI)

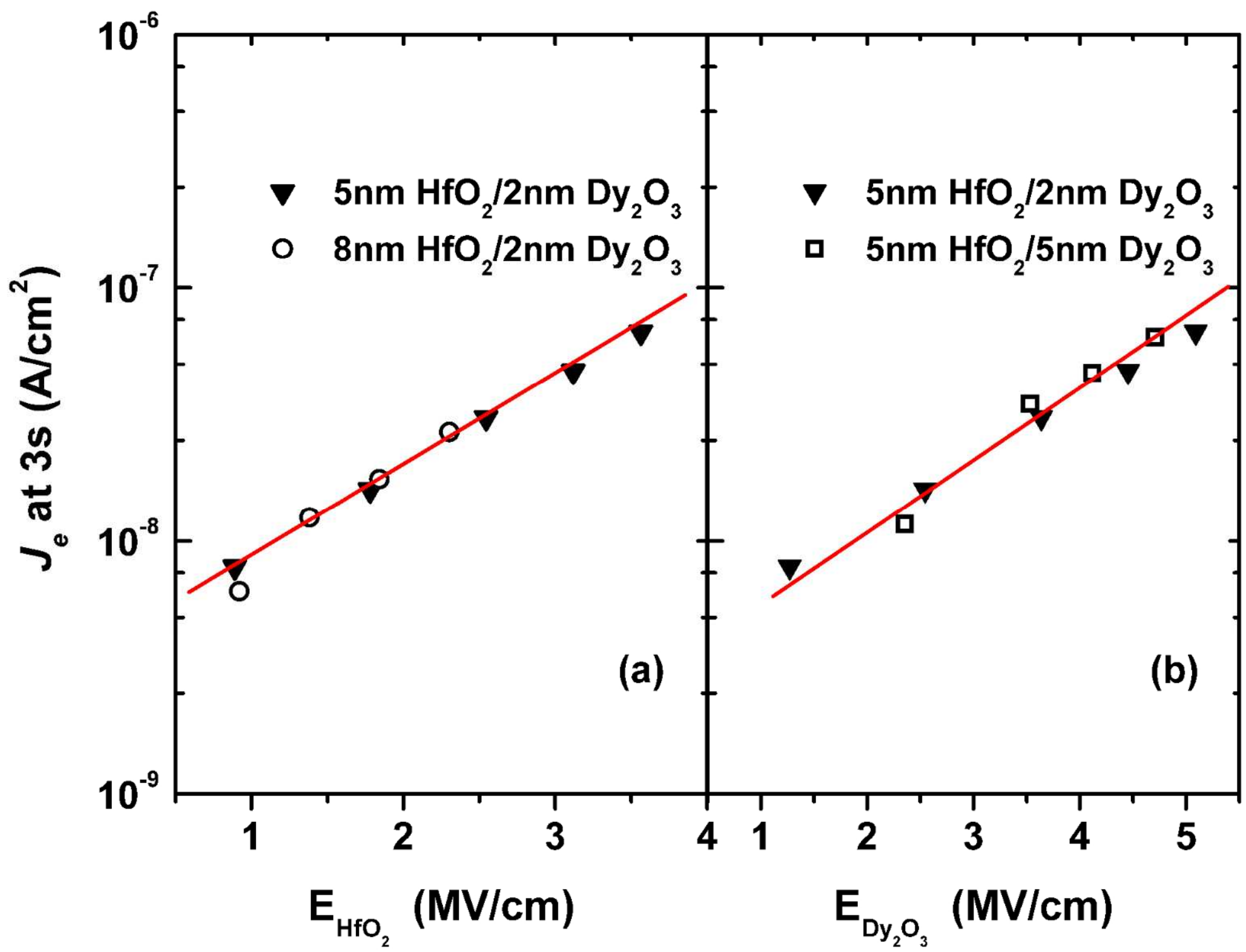

Fig. 6

82x65mm (600 x 600 DPI)

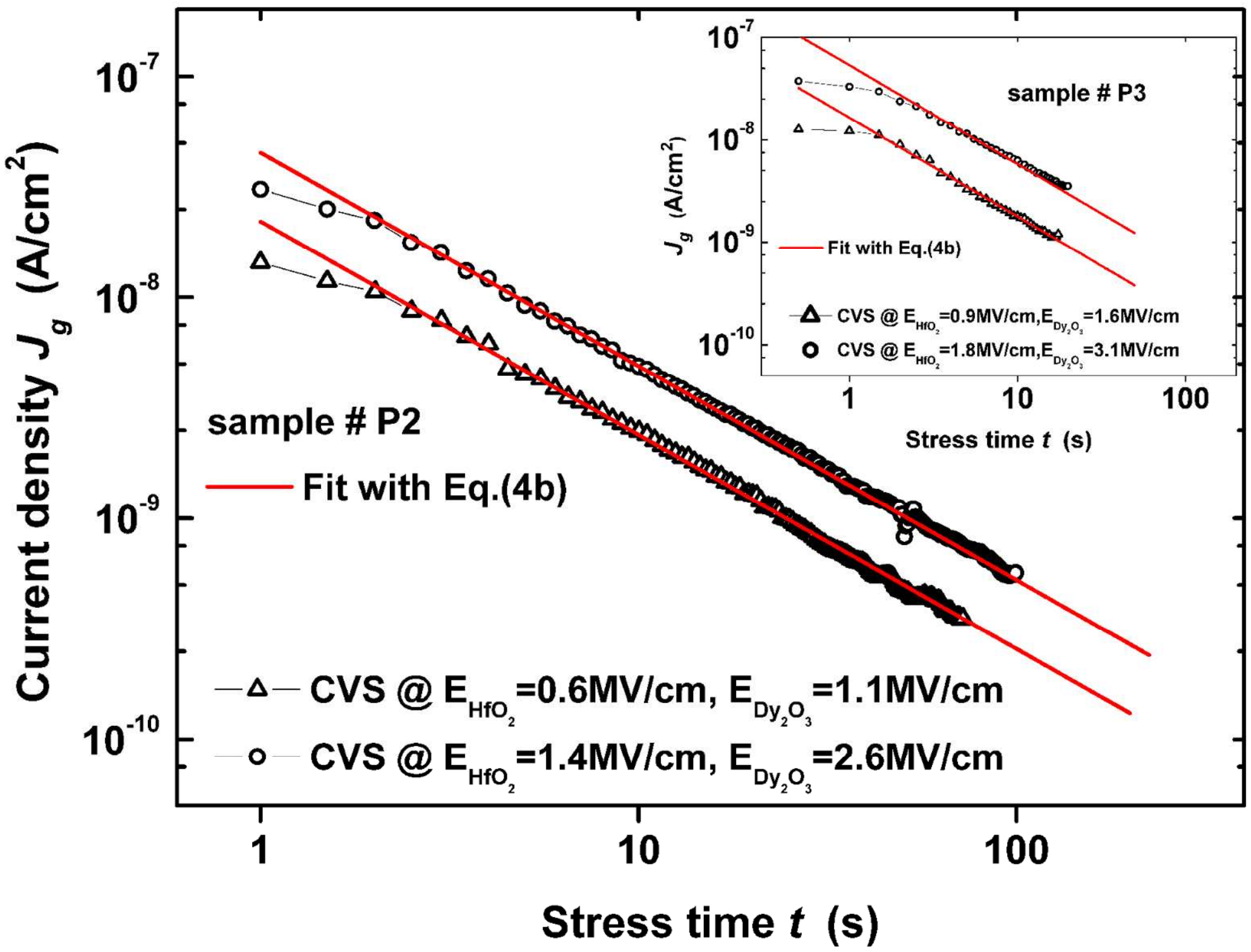

Fig.7

82x65mm (600 x 600 DPI)

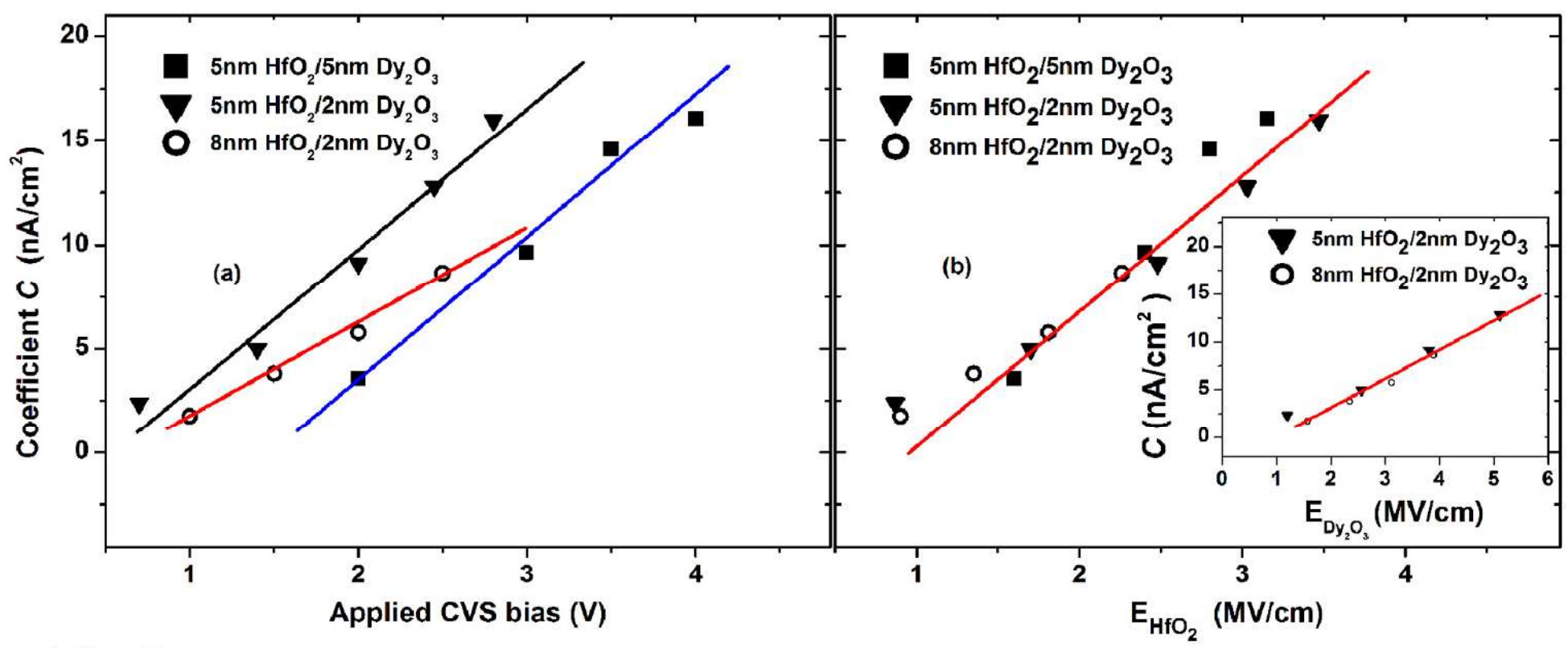

Fig.8

87x39mm (600 x 600 DPI)

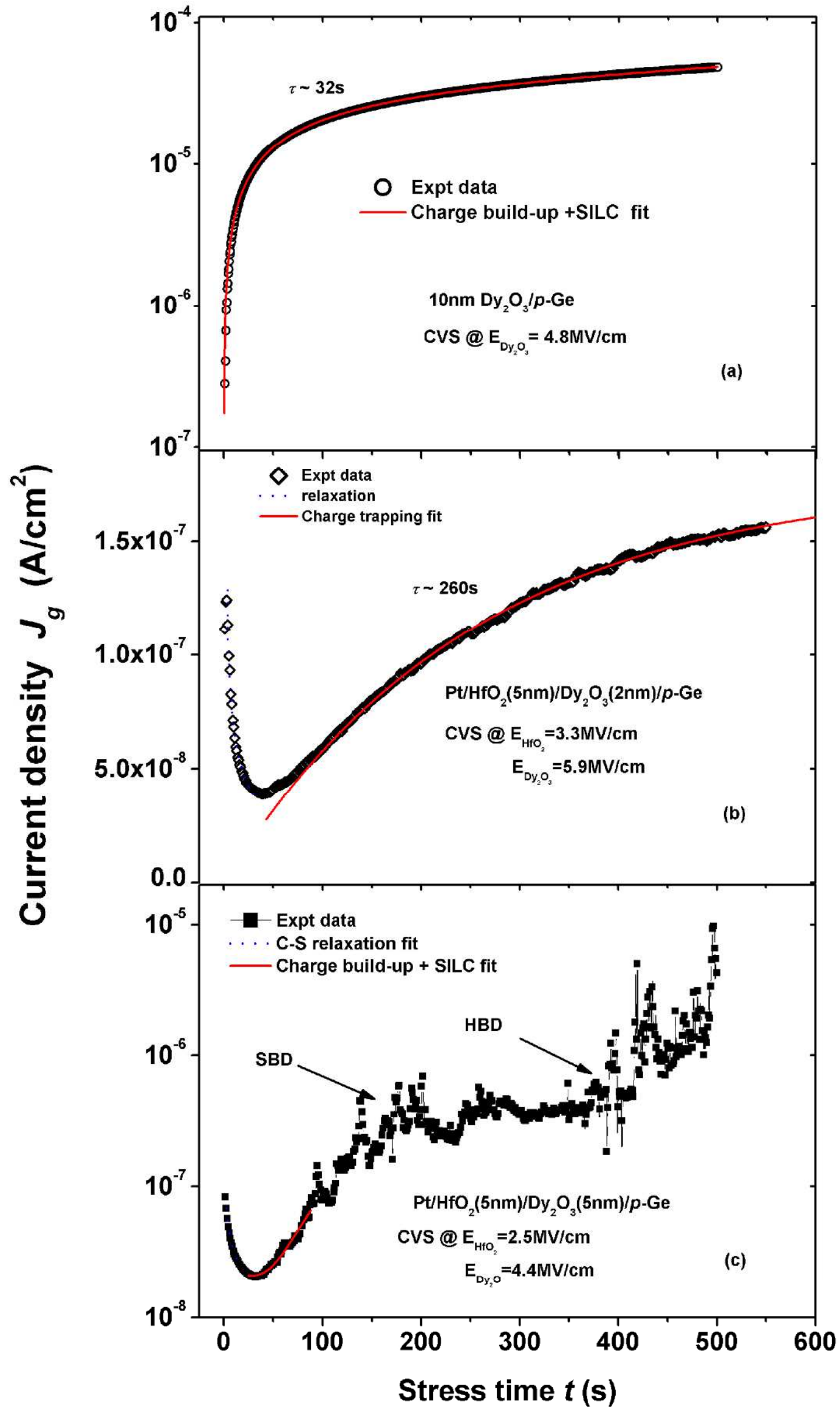

87x140mm (600 x 600 DPI)

**Md. Shahinur Rahman (S'06, M'09)** was born in Narail, Bangladesh. He received B.Sc. (Honours) and M.Sc. degrees in Physics from the University of Dhaka, Bangladesh in 2000, and 2002 respectively. He obtained Ph.D. degree in 2009 from the University of Ioannina, Greece, while his thesis was focused on electrical characteristics and reliability issues of high-κ rare earth oxide gate dielectrics on germanium MOS devices. He was awarded the Greek state PhD fellowship (IKY), postgraduate scholarship of the University of Dhaka and Bangladesh Sena-Kallan scholarship (undergraduate & postgraduate). He is also the recipient of the 'IEEE RS Scholarship 2009' from Reliability Society, IEEE.

He was continuing his research, after his doctoral study, under a Marie Curie Fellowship, MCPAD (Postdoctoral research) of CERN, and was working in the Detector Laboratory of GSI-Helmholtzzentrum für Schwerionenforschung GmbH, Darmstadt, Germany on CVD diamond detectors. Recently he is working as a Semiconductor Radiation Detectors Expert in the OncoRay- Medical Faculty, University of Technology-Dresden, Germany in collaboration with the Radiation Physics department of Helmholtzzentrum Dresden Rossendrof (HZDR). His current research interests include Compound semiconductor (CZT) detectors, Compton camera, Cancer therapy with radiation ion beams, Radiation damage/defects in materials, Detector physics and Particle detectors (e.g. Diamond detectors), also Dielectric/oxide defects and electrical characterization, Reliability issues of CMOS devices, Ge-based MOS devices with high- κ dielectrics, as well as rare earth oxides as high-κ gate dielectrics.

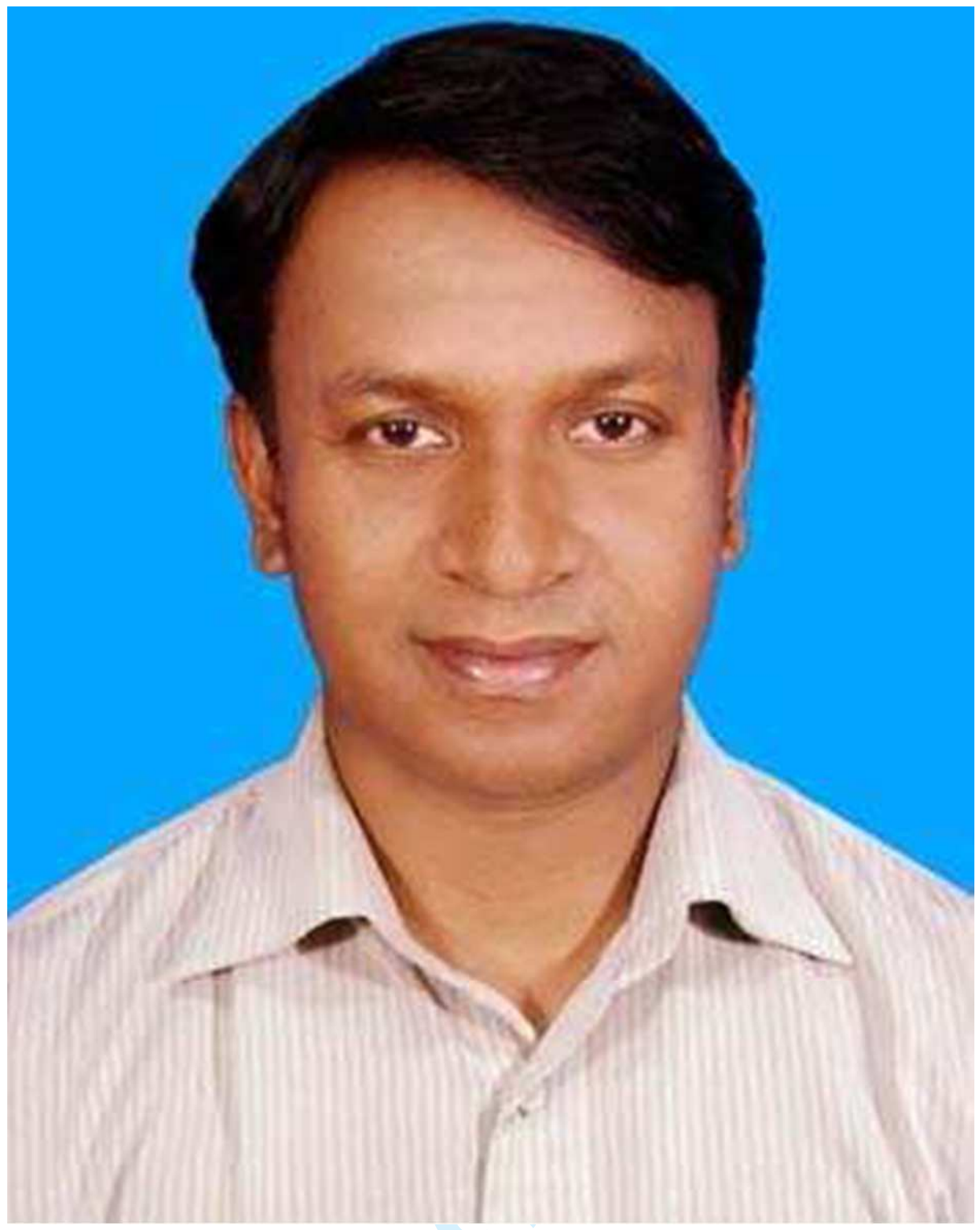

Evangelos Evangelou (M'01) received the B.A. and Ph.D. degrees in Physics from the University of Ioannina, Ioannina, Greece, in 1985 and 1994 respectively for research on defects in III-V based semiconductor devices.

He spent a year as a Research fellow with the Opto-electronic Devices group at the Department of EEE, The Nottingham Trent University, Nottingham, UK, where he worked on the electrical characterization of ACTFEL devices. In 1996 he joined the Department of Physics, University of Ioannina, Ioannina, Greece where he is still working as an Assistant Professor. Since 2000, he participated in several research projects with different industrial partners (ST Microelectronics, IMEC) and research centers (NCSR "Demokritos", MDM-INFM) aiming to study novel high-k materials for potential use in future MOS devices.

His current research interests include the electronic properties of Ge-based MOS devices as well as reliability issues of novel MOS devices with high-k dielectrics.

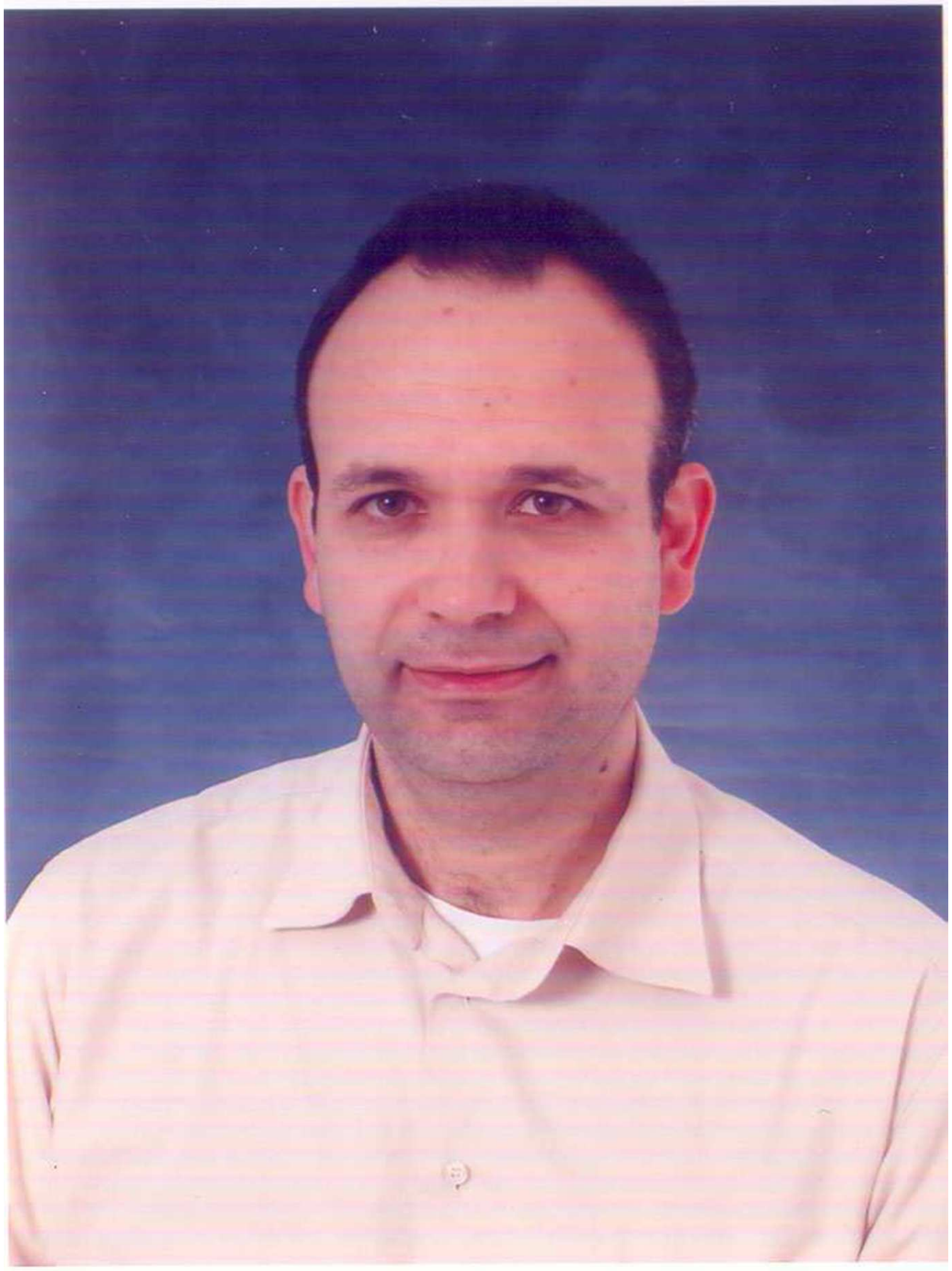